\documentstyle[12pt]{article}
\font\bbexii=cmmib10 scaled\magstep1
\font\bbex=cmmib10
\font\bbevii=cmmib7

\newfam\bbefam
\textfont\bbefam=\bbexii
\scriptfont\bbefam=\bbex
\scriptscriptfont\bbefam=\bbevii
\mathchardef\BigEchar="710F

\begin{document}
\def\lag{\langle}
\def\rag{\rangle}
\begin{titlepage}
\vspace{2.5cm}
\baselineskip 24pt
\begin{center}
\vspace{1.5cm}
\vspace{1.5cm}
\large\bf {Multi-Overlap Simulations}\\
\large\bf {of the $3d$ Edwards-Anderson Ising Spin Glass$^1$}\\
\vspace{1.5cm}
\large{
Bernd A. Berg$^{2,3,4}$ and Wolfhard Janke$^{5,6}$
}
\end{center}
\vspace{2cm}
\begin{center}
{\bf Abstract}
\end{center}

We introduce a novel method for numerical spin glass investigations:
Simulations of two replica at fixed temperature, weighted
such that a broad distribution of the Parisi overlap parameter $q$ is 
achieved. Canonical expectation values for the entire $q$-range 
(multi-overlap) follow by re-weighting. We demonstrate the feasibility 
of the approach by studying the $3d$ Edwards-Anderson Ising
($J_{ik}=\pm 1$) spin glass in the broken phase ($\beta=1$). For the first 
time it becomes possible to obtain reliable results about spin glass
tunneling barriers. In addition, as do some earlier numerical studies, 
our results support that Parisi mean field theory is valid down to $3d$. 
\medskip

\noindent
PACS. 75.40.Mg Numerical simulation studies, 75.50.Lk Spin glasses
and other random magnets
\vfill

\footnotetext[1]{{This research was partially funded by the
Department of Energy under contract DE-FG05-87ER40319.}}
\footnotetext[2]{{Department of Physics, The Florida State University,
                      Tallahassee, FL~32306, USA. }}
\footnotetext[3]{{Supercomputer Computations Research Institute,
                      Tallahassee, FL~32306, USA.}}
\footnotetext[4]{{E-mail: berg@hep.fsu.edu}}
\footnotetext[5]{{Institut f\"ur Physik, Johannes Gutenberg-Universit\"at,
D-55099 Mainz, Germany.}}
\footnotetext[6]{{E-mail: janke@miro.physik.uni-mainz.de}}
\end{titlepage}

\baselineskip 24pt

One of the questions which ought to be addressed before performing a
large scale computer simulation is ``What are suitable weight factors
for the problem at hand?'' The weight factor of canonical Monte Carlo (MC)
simulations is $\exp (-\beta E)$, where $E$ is the energy of the 
configuration to be updated and $\beta$ is the inverse
temperature in natural units. The Metropolis and other methods generate
canonical configurations through a Markov process.
It has been expert wisdom~\cite{ToVa77} for quite a while and became 
widely recognized in recent years
\cite{BeNe92,MaPa92,BeHaNe93,Many,Be97,Ja97} 
that MC simulations with
a-priori unknown weight factors, like for instance the inverse
spectral density $1/n(E)$, are also feasible and deserve to be considered. 
But it is not straightforward to design suitable weights. To find a 
weighting procedure that works in practice requires considerable intuitive 
or other understanding of the underlying physics.

On the easier side are situations where one has to enhance rare
configurations which are controlled by some standard thermodynamic
observable.
They are called {\it static} in Ref.~\cite{Be97}. An example are
temperature-driven first-order phase transitions, controlled
by the energy. The multicanonical method~\cite{BeNe92} defines
energy dependent weight factors to enhance configurations needed to
estimate the interfacial tension. The multimagnetical 
method~\cite{BeHaNe93} does the same for magnetic field-driven 
first-order transitions, controlled by the magnetization.
Far more difficult are problems, called {\it dynamic} in Ref.~\cite{Be97}, 
for which the ergodicity of canonical MC simulations breaks down due 
to energy barriers for which no 
explicit parameterization in terms of a standard
thermodynamic variable is known. This is the case for spin glasses and
other systems with conflicting constraints where the energy barriers
are caused by disorder and frustration. Below some freezing temperature 
it becomes extremely difficult to generate canonical equilibrium 
configurations for these systems. Recently, progress has been made by
exploring~\cite{BeCe92,KeRe94,HuNe96,MaPaRu97} innovative weighting
methods for this problem. Along this line our paper introduces a novel,
efficient approach.

We focus on the $3d$ Edwards-Anderson Ising (EAI) spin glass on a simple
cubic lattice. It is widely considered to be the simplest model to 
exhibit realistic spin glass behavior and has been the testing ground 
of Refs.~\cite{BeCe92,KeRe94,HuNe96,MaPaRu97}. The energy is given by
\begin{equation} \label{energy}
E = - \sum_{\langle ik \rangle} J_{ik}\, s_i s_k \qquad ,
\end{equation}
where the sum is over nearest-neighbour sites. The Ising spins $s_i$
and $s_k$ as well as the exchange coupling constants $J_{ik}$ take
values~$\pm 1$. A realization is defined by a fixed assignment of the 
exchange coupling constants $J_{ik}$. In our investigation we
enforce the constraint $\sum_{\langle ik \rangle}J_{ik}=0$
by picking half of the $J_{ik}$ at random and assigning them the
value $+1$, whereas the others are fixed at $-1$. 

Early MC simulations of the EAI model, for a concise review
see~\cite{MaPaRu97}, located the freezing temperature at 
$\beta_c \approx 0.9$. Recent, very high statistics canonical
simulations~\cite{KaYo96,remark} estimate $\beta_c=0.901\pm~0.034$, and 
considerably improve the evidence in support of a second-order phase 
transition at $\beta_c$. The
studies~\cite{BeCe92,KeRe94,HuNe96,MaPaRu97} focus on first improving
the notorious slowing down of simulations at $\beta > \beta_c$. Their main
idea is to avoid getting stuck in metastable low-energy states by
using a Markov process which samples the ordered as well as the
disordered regions of configuration space in one run. Ref.~\cite{BeCe92}
does this by multicanonical~\cite{BeNe92} re-weighting in the energy, 
whereas Refs.~\cite{KeRe94,HuNe96,MaPaRu97} use and improve the method
of enlarged ensembles~\cite{MaPa92} in their simulated tempering 
version. Refreshing the system in the disordered 
phase clearly benefits the simulations, but the performance has
remained below early expectation. It has been conjectured that the reason 
lies in the tree-like structure of the low-energy spin glass states, 
see Ref.~\cite{BhSe97} for a detailed discussion.

Studying simulated tempering, Kerler and Rehberg~\cite{KeRe94} combined 
two copies (replica) of the same realization (defined by its couplings
$J_{ik}$) in one simulation. The purpose was to allow for direct 
evaluation of the Parisi overlap parameter
\begin{equation} \label{q}
q = {1\over N} \sum_{i=1}^N s_i^1 s_i^2\qquad . 
\end{equation}
Here $N$ denotes the number of spins, the spins $s^1_i=\pm 1$ correspond 
to the first and the spins $s^2_i=\pm 1$ to the second replica. Now, our 
observation is that one does still control canonical expectation values at 
temperature $\beta^{-1}$ when one simulates with a weight function
\begin{equation} \label{weight}
 w(q) = \exp \left[ \beta \sum_{\langle ik \rangle} J_{ik}
(s^1_i s^1_k + s^2_i s^2_k) + S(q) \right] \qquad ,
\end{equation}
This is obvious for $S(q)=0$, and a non-trivial $S(q)$ can be mapped 
onto this situation by standard re-weighting~\cite{ToVa77,BeNe92}. Of
particular interest is to determine $S(q)$ recursively~\cite{Be96}
such that the histogram $H(q)$ becomes uniform in $q$ and the 
interpretation of $S(q)$ being the microcanonical entropy of the 
Parisi order parameter. Hence, although an explicit order parameter does
not 
exist, an approach very similar to the multimagnetical~\cite{BeHaNe93} 
(which is an highly efficient way to sample interface barriers for 
ferromagnets) exists herewith. In contrast to this multi-overlap 
method, simulated and 
parallel tempering techniques~\cite{KeRe94,HuNe96,MaPaRu97} do not 
allow to change barrier heights.

Our EAI simulations are performed on $V=L^3$ lattices at $\beta =1$, 
in the interesting region well below the freezing temperature. All
calculations were done on a cluster of Alpha workstations at FSU. We
simulated 512 different realizations for $L=4,\, 6$, and 8, and 33 for
$L=12$. 
For all realizations
tunneling between the extrema $q=\pm 1$ was achieved. Each production
run of data taking was concluded after at least twenty
tunneling event of the form 
$$ (q=0) \to (q=\pm 1) ~~{\rm and\ back}$$
were recorded (for technical reasons the actual numbers were between 
20 and 39 per realization). Table~1 gives an overview of the tunneling 
performance of our algorithm. Fitting the estimates of the 
mean value $\overline{\tau}$ to the form
$ \ln (\overline{\tau}) = a + z\, \ln (V) $ gives 
$z=2.42\pm 0.03$. Compared with the slowing down of \cite{BeCe92} this 
is an improvement of almost a factor $\sqrt{V}$. Still, the slowing 
down is far off from the theoretical optimum \cite{BeNe92} $z=1$. One
reason seems to be that we are enforcing the limit $q\to\pm 1$. This
limit correlates strongly with ground states, which are difficult to 
reach
by local updates, see for instance \cite{BhSe97}. Being content with a 
smaller region (like the two outmost maxima in the $q$-distribution) is 
expected to give further improvements of the tunneling performance.
Other data compiled in Table~1 are the encountered minimum, maximum and
median tunneling times. We observe that the mean values are 
systematically larger than the median, what means that the tunneling 
distribution has a rather long tail towards large tunneling times.
On the other hand, the effect is not severely hindering our
multi-overlap simulations: For the lattice sizes $L=4$ to 8 the
worst behaved realization took never more than 3\% of the entire
computer time and for $L=12$ (where we have only 33 realizations)
this amount was 12\%.

Initially in each run, a working estimate of the weight 
function~(\ref{weight}) has to be obtained. Using a variant of the
recursion proposed in~\cite{Be96} this has turned out to be remarkably
easy. For each case we stopped the recursion of weights after four
tunnelings were achieved and the used computer time corresponds in 
good approximation to $4\overline{\tau}$, with $\overline{\tau}$ as
given in Table~1.

The analysis of the thus created data allows us to calculate a number of 
physically interesting quantities. In particular accurate determinations 
of the canonical potential barriers in $q$ are, for the first time, 
possible. Let $P_i (q)$ be the canonical probability densities of $q$,
where $i=1,...,n$ labels the different realizations (additional 
dependence on lattice size and temperature is implicit). We define the 
corresponding potential barrier by
\begin{equation} \label{barrier}
B_i = \prod_{q=-1}^{-\, \triangle q} \max \left[ 1, 
      P_i(q)/P_i(q+\triangle q) \right]\, ,
\end{equation}
where $\triangle q$ is the stepsize in $q$. For the double-peak
situations of first-order phase transitions \cite{BeHaNe93}
Eq.~(\ref{barrier}) becomes $B_i=P_i^{\max}/P_i^{\min}$,
where $P_i^{\max}$ is the absolute maximum and $P_i^{\min}$ is the
absolute minimum (for ferromagnets at $q=0$) of the probability
density $P_i(q)$. Our definition generalizes to the situation where
several minima and maxima occur due to disorder and frustration.
When evaluating (\ref{barrier}) from numerical data for $P_i (q)$
some care is needed to avoid contributions from statistical 
fluctuations of $P_i(q)$.

Graphically, our values for the $B_i$ are presented in Fig.~1. 
It comes as a surprise that the finite-size dependence of the 
distributions is very weak. Therefore, one may question the 
apparently accepted opinion that these barriers are primarily 
responsible for the severe slowing down of canonical MC simulations 
with increasing lattice size. To study this issue further, we have 
compiled in Table~2 for each lattice size the following informations 
about our potential barrier results: largest and second largest 
values $B_{\rm max}$ and $B_2$, median values and 70\% confidence 
limits around those, and mean 
values $\overline{B}$ with statistical error bars. From this table it
becomes
obvious, why this investigation could not be performed using
canonical methods to which also enlarged ensembles belong (they
enlarge the ensemble but still use canonical weights). For these methods 
the slowing down would be proportional to the average barrier height
$\overline{B}$, which
is already large for $L=4$, about 18 thousand, and increases to 
about 2.8 million for $L=12$. 

Next, the reader may be puzzled by the very large error bars assigned
to the mean barrier values. Their explanation is: The entire 
mean value is dominated by the largest barrier, which contributes
between 70\% $(L=4)$ and, practically, 100\% $(L=12)$, see $B_{\max}$ 
in the second column of Table~2. Besides $B_{\max}$, the second largest 
value $B_2$ is listed in the third column. The lesson from these numbers 
is that very few of the realizations are responsible for the collapse of 
canonical simulation methods. It may be remarked that most of these worst 
case barriers exhibit simple double-peak behavior. An exception is
$L=6$ where the distribution yielding $B_{\max}$ has two double peaks. 
To exhibit the difference, the inlay of Fig.~1 depicts the 
right-hand-side of the $L=6$ probability densities $P_i(q)$ with 
$i=459$ corresponding to the $B_{\max}$ and $i=122$ to the $B_2$
barrier. For $B_2$ the value of our barrier definition (\ref{barrier})
agrees with the $P_{\max}/P_{\min}$ value, whereas for $B_{\max}$ it is 
by about a factor of two larger. 

Typical configurations, described by the  median results of Table~2, 
have much smaller tunneling barriers. They turn out to be quite
insensitive to the lattice size, in fact the value 
$B_{\rm med}=12.3$ fits into the confidence interval for all simulated 
lattice sizes. Presumably, there is some increase of $B_{\rm med}$ with 
lattice size, but to trace it we would need to simulate more 
realizations. This result of an almost constant typical 
tunneling barrier is consistent with the fact
that our tunneling times are rather far apart from their
theoretical optimum: Other reasons than overlap barriers have to
be responsible.

Our data are consistent with other numerical 
evidence \cite{KaYo96} in favor of the Parisi mean field scenario 
being valid down to $3d$ and against the competing droplet picture, 
thus indirectly supporting Parisi's criticism \cite{Pa96} 
of Ref.~\cite{NeSt96}. The averaged canonical
probability densities $P(q) = [P_i(q)]_{\rm av}$ at the
simulation point $\beta = 1.0$ are shown in Fig.~2. While the peak moves
with increasing lattice size towards smaller $q$-values, the value 
of $P(0)$ is clearly non-zero and shows almost no finite-size dependence.
The $\beta$-dependence of $P(q)$ obtained by standard reweighting of
our time-series data at $\beta=1$ is illustrated in Fig.~3 for $L=8$. 
For lower temperatures the peak of $P(q)$ becomes more pronounced and
moves towards larger $q$-values.
The extremal $\beta$-values indicate
the inverse temperature range in which reliable results can be 
expected. This range was estimated by measuring the overlap of
the reweighted energy histogram with the energy histogram at the 
simulation point $\beta = 1.0$, individually for each of the
realizations. With the present statistics the phase transition point
should thus be in this range, at least up to $L=8$. By analyzing the
spin glass susceptibility, $\chi_{\rm SG} = N [\langle q^2 \rangle
]_{\rm av}$, we obtain the best finite-size scaling fit
$\chi_{\rm SG} \propto L^{\gamma/\nu}$ at $\beta_c = 0.88$ with $\gamma/\nu
= 2.37(4)$ and a goodness-of-fit parameter $Q=0.25$. This is corroborated
by the curves of the Binder parameter, $g = (1/2) (3 - [\langle q^4
\rangle]_{\rm av}/[\langle q^2 \rangle ]_{\rm av}^2)$, which merge 
around $\beta=0.89$. In the
low-temperature phase ($\beta > \beta_c$) the curves for different
lattice sizes seem to fall on top of each other, but our error bars 
are still too large to draw a firm conclusion from this quantity.
These results are consistent with the findings of Ref.~\cite{KaYo96} 
and could be easily improved by redoing the simulations closer to 
$\beta_c$, possibly with more realizations and less statistics per 
realization. For such,
or similar, studies the number of $L=12$ realizations can be readily 
enhanced by running on a parallel computer like a Cray T3E. Narrowing
the $q$-range will allow to simulate lattices of size $L=16$ and
beyond.

Finally, we like to mention that our method is particularly well-suited
to study the influence of an interaction term~\cite{CaPa90}
$$ \epsilon \sum_{i=1}^N s^1_i\, s^2_i = \epsilon\, N\, q $$
in the Hamiltonian (\ref{energy}): We obtain expectation values for 
arbitrary $\epsilon$-values. Physically most interesting is to combine 
a non-zero magnetic field with a non-zero $\epsilon$-value.

In conclusion, we have demonstrated the feasibility of using 
$q$-dependent (multi-overlap) weight factors. Although the tunneling 
performance is not optimal, the method opens new horizons for spin 
glass simulations. In this paper we succeeded, for the first time,
to study $q$-barriers in some details. Using parallel computers and
slight modifications of our method (like narrowing the $q$-range,
including a magnetic field, etc.) will allow to extend our investigation
into various interesting directions, like an improved study of the
thermodynamic limit at and below the transition point, or 
$\epsilon$-physics.
\bigskip

\noindent {\bf Acknowledgements:}
W.J. acknowledges support from the Deutsche Forschungsgemeinschaft
(DFG) through a Heisenberg Fellowship, and funding through the US Department
of Energy enabled his visit at the Florida State University. Major parts
of this paper were completed when both authors participated in the
research group {\it Multi-Scale Phenomena} at the ZIF of Bielefeld
University. We like to thank the organizers, in particular Frithjof 
Karsch, for their hospitality. 
\hfill\break

\hfill\break 

\section*{Tables and Figure Captions}

\begin{table}[ht]
\centering
\begin{tabular}{||c|c|c|c|c||}                    \hline
$L$& $\tau_{\min}$& $\tau_{\max}$& $\tau_{\rm med}$& 
					      $\overline{\tau}$  \\ \hline
 4 &     4.5E02  &     6.2E03  &     9.9E02  & (1.13$\pm$0.03)E03 \\ \hline
 6 &     4.9E03  &     3.1E05  &     1.3E04  & (1.88$\pm$0.09)E04 \\ \hline
 8 &     2.4E04  &     1.6E06  &     1.1E05  & (1.76$\pm$0.09)E05 \\ \hline
12 &     7.1E05  &     1.6E07  &     2.7E06  & (4.11$\pm$0.65)E06 \\ \hline
\end{tabular}
\caption{{\em Overview of the tunneling performance: minimum, maximum,
median
and mean $\pm$ error tunneling times. All numbers are in units of sweeps.}}
\end{table}

\begin{table}[ht]
\centering
\begin{tabular}{||c|c|c|c|c|c|c||}                               \hline
$L$& $B_{\max}$     & $B_2$ &$B^+_{\rm med}$ &
		         $B_{\rm med}$&$B^-_{\rm med}$&${\overline B}$\\ \hline
 4 & 6.56E06 (70\%) &9.11E05& 15.1 & 12.4 & 9.62 & $(1.84\pm 1.30)$E04\\
\hline
 6 & 2.76E06 (74\%) &1.44E05& 12.3 & 11.1 & 10.1 & $(7.29\pm 5.42)$E03\\
\hline
 8 & 1.97E08 (98\%) &1.36E06& 17.7 & 15.2 & 12.3 & $(3.91\pm 3.85)$E05\\
\hline
12 & 9.14E07 (100\%)&1.96E03& 35.3 & 12.9 & 10.7 & $(2.77\pm 2.77)$E06\\
\hline
\end{tabular}
\caption{{\em Canonical potential barriers: maximum (and its contribution
to the
mean in \%), second largest value, upper median confidence limit, median,
lower
median confidence limit (upper and lower limit bound a 70\% confidence
interval),
the mean and its error bar.}}
\end{table}


\begin{description}

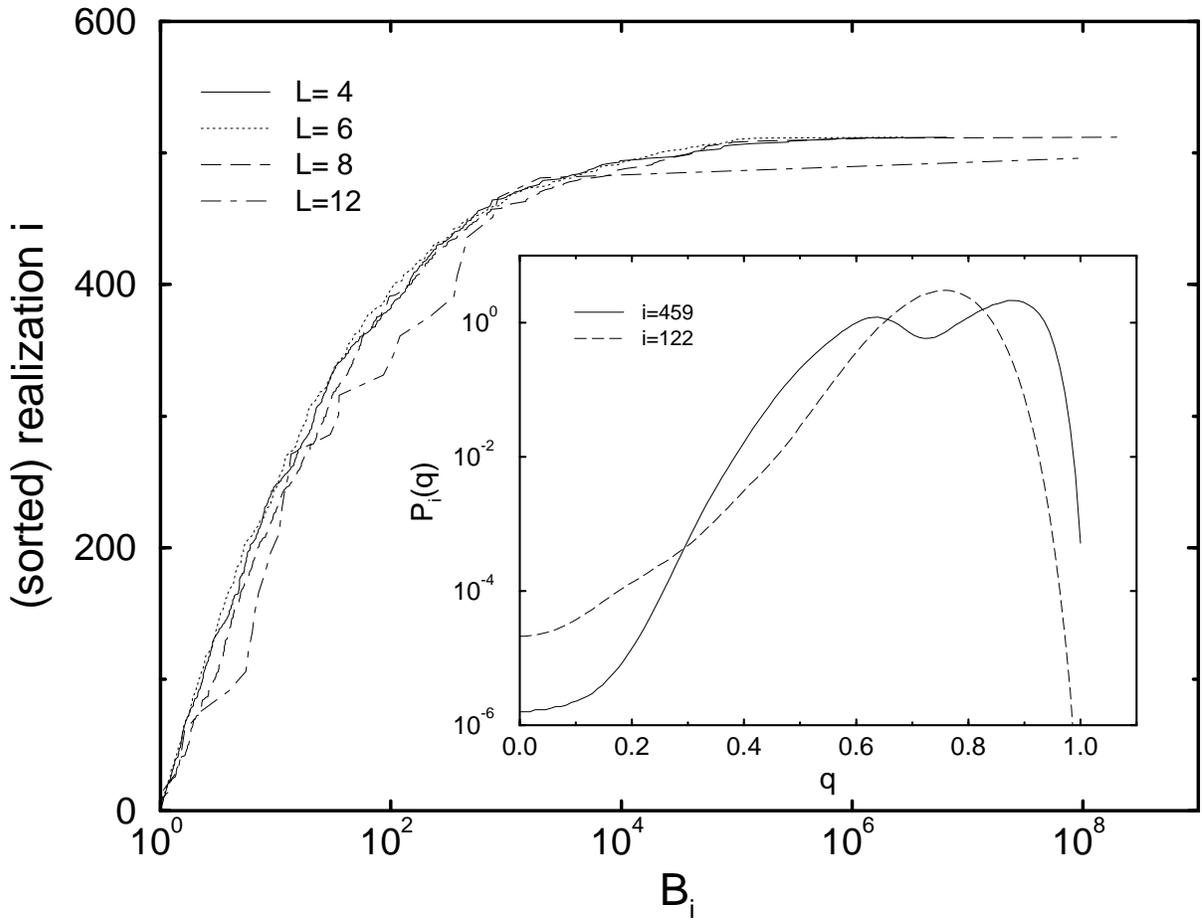
\item{Figure 1:}
{\it Canonical tunneling barrier distributions at $\beta = 1$.
(The $L=12$ barriers are re-labelled to fill into the 1--512 range.)
The inlay shows the two worst $L=6$ realizations. }

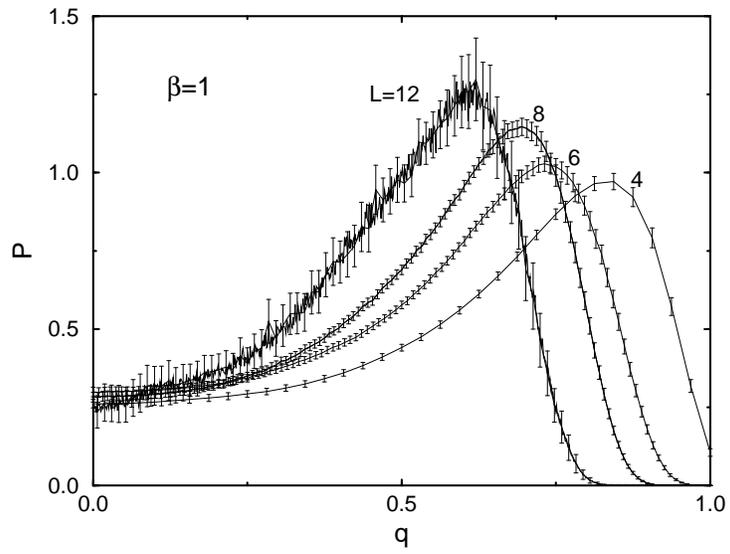
\item{Figure 2:}
{\it Finite-size dependence of the
averaged canonical probability densities $P(q)$ 
at $\beta = 1$. For $L=8$ only every second error bar is shown, and 
for $L=12$ only every tenth.}

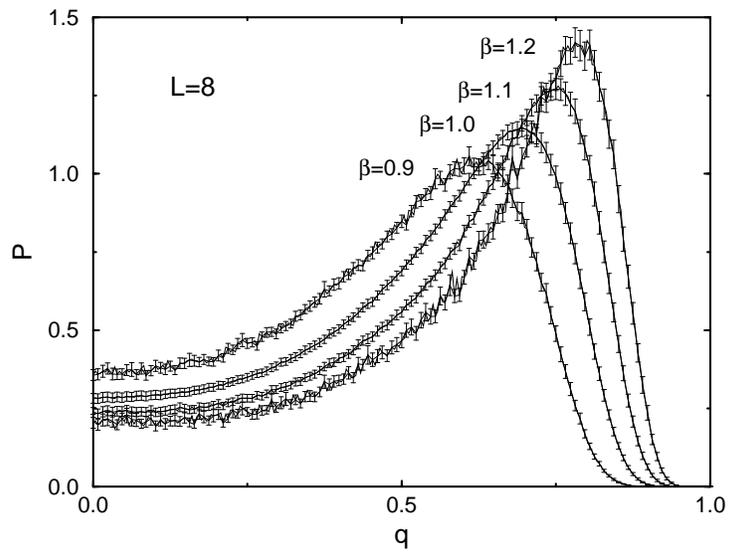
\item{Figure 3:}
{\it Temperature dependence of the
averaged canonical probability densities $P(q)$ 
for $L=8$, obtained by reweighting. Only every second error 
bar is shown.}

\end{description}





\begin{thebibliography}{12}
\bibitem{remark} In this study the realizations are generated according 
                 to the condition that $J_{ik}$ takes the values 
                 $\pm 1$ with equal probability, which is weaker than
                 $\sum_{\langle ik\rangle} J_{ik}=0$. The thermodynamic
                 limits are expected to agree.
\end{thebibliography}
\end{document}